\documentclass[10pt,onecolumn]{article}
\usepackage{amsmath}
\usepackage{amsfonts}
\usepackage{amssymb}

\usepackage[latin1]{inputenc}
\usepackage{latexsym} 
\usepackage{graphicx}
\usepackage{psfrag}
\usepackage{exscale}
\usepackage[permil]{overpic}       
\usepackage[vflt]{floatflt}        
\usepackage{times}
\usepackage{makeidx}               
\usepackage{color}
\usepackage{framed,color}

\usepackage{hyperref}
\usepackage[numbers,sort&compress]{natbib}
\usepackage{hypernat}
\usepackage{wasysym}
\usepackage{rotating}

\makeindex   

   {\begin{list}{}{
      \settowidth{\labelwidth}{\textbf{#1}}
      \setlength{\leftmargin}{\labelwidth}
         \addtolength{\leftmargin}{\labelsep}
      \setlength{\parsep}{0.5ex plus0.2ex minus0.2ex}
      \setlength{\itemsep}{0.3ex}
      }}
   {\end{list}}

\newcommand{\ze}[2]{#1\;{\rm#2}} 

\newfont{\tensy}{cmsy10}           

\newcommand{\fm}[1] 
{\begin{displaymath}#1\end{displaymath}}

\newcommand{\nfm}[1] 
{\begin{equation}#1\end{equation}}

\newcommand{\nmzfm}[1] 
{\begin{eqnarray}#1\end{eqnarray}}

\newcommand{\mzfm}[1] 
{\begin{eqnarray*}#1\end{eqnarray*}}

\newcommand{\gfm}[1] 
{\begin{displaymath}\fbox{$\displaystyle #1$}\end{displaymath}}

\newcommand{\ngfm}[1] 
{\begin{equation}\fbox{$\displaystyle #1$}\end{equation}}

\newcommand{\mzgfm}[1] 
{\begin{displaymath}\fbox{$\begin{array}{rcl} #1\end{array}$}\end{displaymath}}

\newcommand{\figcapt}[2] 
{\renewcommand{\normalsize}{\small}\textbf{\caption[#1]{\textmd{#2}}}}

\newcommand{\fif}[1] 
{\mathbf{#1}}

\newcommand{\bi}[4] 
{\int_{#1}^{#2}{#3}\,{#4}}

\newcommand{\ubi}[2] 
{\int{#1}\,{#2}}

\definecolor{shadecolor}{gray}{.90} 

\newtheorem{beispiels}{Beispiel}[section] 

\newtheorem{aufgabeS}{Aufgabe}[section]

\newcommand{\qM} 
{\left[\ddots\right]}

\newcommand{\SV} 
{\left[\vdots\right]}

\title{High-Intensity Discharge Lamp and Duffing Oscillator - Similarities and Differences}

\author{\normalsize Bernd Baumann$^{*1}$,  Joerg Schwieger$^{1, 2}$, Ulrich Stein$^{1}$ \\ \normalsize Sarah Hallerberg$^{1}$ and Marcus Wolff$^{1}$}

\date{\footnotesize
$^{1}$Hamburg University of Applied Sciences,  Department of Mechanical Engineering and Production,\hfill \phantom{.}\\
\hspace{1mm} Berliner Tor 21, 20099 Hamburg,  Germany\hfill \phantom{.}\\
$^{2}$University of the West of Scotland, Paisley Campus, School of Engineering and Computing, High Street, \hfill \phantom{.}\\
\hspace{1mm} Paisley PA1 2BE, United Kingdom \hfill\phantom{.}\\
$^*$ info@BerndBaumann.de\hfill\phantom{.}}

\begin{document}
\pdfoutput=1
\maketitle

\begin{abstract}
The processes inside the arc tube of high-intensity discharge lamps are investigated by finite element simulations. The behavior of the gas mixture inside the arc tube is governed by differential equations describing mass, energy and charge conservation as well as the Helmholtz equation for the acoustic pressure and the Navier-Stokes equation for the flow driven by the buoyancy and the acoustic streaming force. The model is highly nonlinear and requires a recursion procedure to account for the impact of acoustic streaming on the temperature and other fields.
The investigations reveal the presence of a hysteresis and the corresponding jump phenomenon, quite similar to a Duffing oscillator. The similarities and, in particular, the differences of the nonlinear behavior of the high-intensity discharge lamp to that of a Duffing oscillator are discussed. For large amplitudes the high-intensity discharge lamp exhibits a stiffening effect in contrast to the Duffing oscillator.

\end{abstract}

\section{Introduction}
\begin{figure}
\centering
\includegraphics[width=0.7\linewidth]{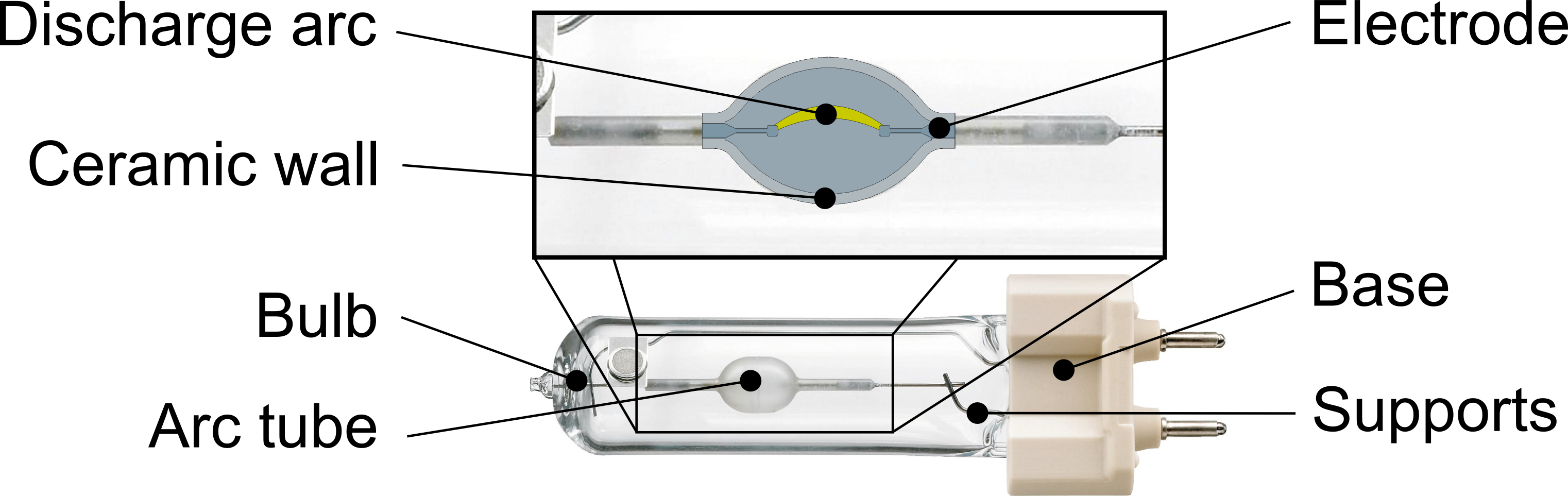}
\caption[Kurze caption]{Design of a MH lamp (Philips 35 W 930 Elite). The arc tube is filled with argon, mercury and metal halides. The distance between the electrodes is $\ze{4.8}{mm}$. The upward bending of the plasma arc is due to buoyancy. The temperature in the plasma arc is about $\ze{6000}{K}$ and drops to ca.~$\ze{1500}{K}$ at the ceramic walls.}
\label{fig:hidlamp}
\end{figure}

High-intensity discharge (HID) lamps are encountered in many lighting applications. State of the art are metal halide (MH) lamps (Figure~\ref{fig:hidlamp}), which possess a sun-like light characteristic.
To avoid a demixing of the arc tube content and to reduce erosion of the electrodes, HID~lamps are typically operated at $\ze{400}{Hz}$ alternating current. The size and the cost of the drivers could be reduced considerably, if the lamps would be operated at approximately $\ze{300}{kHz}$~\cite{Trestman.2002}. Unfortunately, at these high frequencies the periodic heating of the plasma induces acoustic resonances, which, in turn, result in light flicker and even severer problems~\cite{Flesch.2006}. In previous articles we have presented results of  finite-element (FE) simulations, which have been carried out to obtain an understanding of the mechanisms responsible for light flicker \cite{Baumann.2009, BaumannEtAl.2015,0022-3727-49-25-255201}. Here we elaborate on the interpretation of the findings in terms of a nonlinear oscillator.

\section{Modeling}
Dreeben pursued the most direct approach for the modeling of an HID~lamp by setting up a transient model~\cite{Dreeben.2008}. In order to obtain results within reasonable CPU-time, he had to restrict himself to a 2D-model corresponding to a lamp with an arc tube of infinite length. To attain a more realistic description, we use a 3D-model. We set up a time-independent model in the frequency domain to compensate for the additional degrees of freedom. The idea is to identify light flicker by the appearance of instable solutions. It is not possible to take advantage of the axisymmetric lamp design, since for a horizontally operated lamp the upward bending of the plasma arc due to buoyancy violates the symmetry.

A relatively simple model describing the processes in the ceramic walls, the electrodes and the interior of the arc tube comprises coupled partial differential equations (PDE) representing the conservation of mass, energy, electric charge and momentum, respectively (thermal plasma model~\cite{Flesch.2006}). The Navier-Stokes equation (momentum conservation) has to be supplemented by the equation of state of an ideal gas. To model the generation and propagation of sound waves, three additional PDEs for the acoustic pressure, the acoustic temperature and the sound particle velocity are needed (visco-thermal acoustic model \cite{Joly_2010}). It has been suspected that acoustic streaming (AS) plays an important role for the generation of light flicker~\cite{Afshar.2008}. AS is a directed flow in a fluid, driven by high amplitude periodic sound waves. It is incorporated into the model by a force term in the Navier-Stokes equation. The model is completed by appropriate boundary conditions and coefficients describing the physical properties of the materials. Some of these coefficients such as viscosity and thermal conductivity are temperature dependent and, therefore, field variables themselves. Their temperature dependencies are described by heuristic functions~\cite{BaumannEtAl.2015}. 

\begin{figure}
\centering
\includegraphics[width=0.7\linewidth]{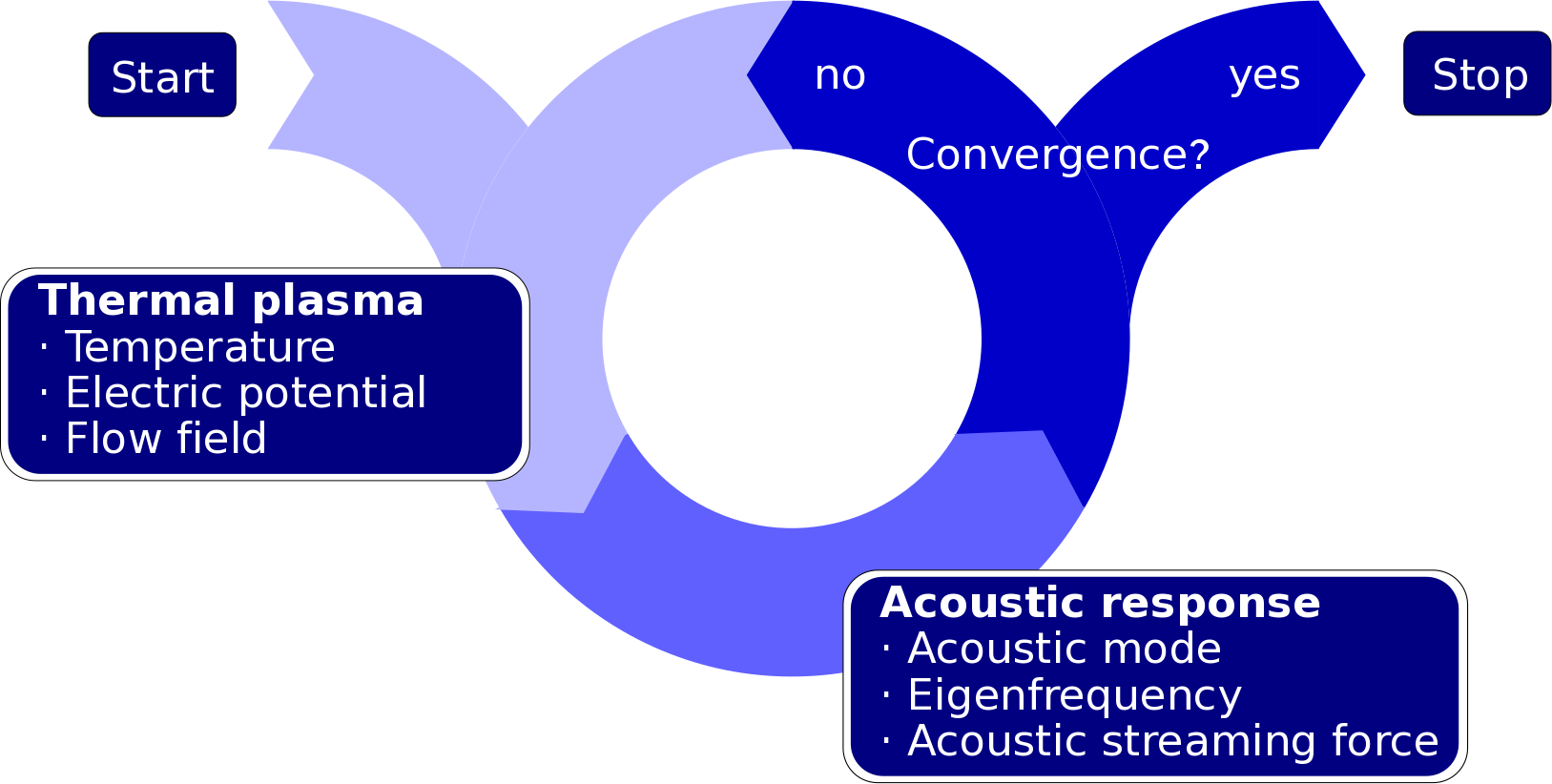}
\caption{Schematic description of the recursion.}
\label{fig:recursion3}
\end{figure}

Due to the complexity of this model, it is not suitable for the computer resources at our disposal.
A considerable reduction of the required computer resources can be obtained by replacing the three PDEs for the modeling of sound waves by the inhomogeneous Helmholtz equation for the acoustic pressure. The inhomogeneity describing the generation of sound is identical to the resistive  loss density minus the power density for the generation of electromagnetic radiation. The solution of the Helmholtz equation can be expressed in terms of an acoustic mode expansion, and the modes can readily be obtained numerically. This approach has the disadvantage of not taking acoustic loss into account. However, it is possible to incorporate loss by means of loss factors~\cite{Kreuzer.1977, Baumann.2007, BaumannEtAl.2015}. In our model, we have included viscous loss and loss due to thermal conductivity in the interior of the arc tube and at the inner surface of the arc tube as well.
The simulations based on the mode expansion model consist of several consecutive steps (Figure~\ref{fig:recursion3}):

\begin{enumerate}
\item The Thermal Plasma Model is solved. The main results of this step are the temperature field and the inhomogeneity term of the Helmholtz equation.

\item On the basis of the temperature field, the acoustic modes and natural frequencies are determined. The inhomogeneity term and the loss factors allow determining the amplitude of the sound pressure at a certain excitation frequency.

\item The sound particle velocity is calculated from the acoustic pressure. It has to be multiplied by a heuristical factor, which is equal to one in the interior of the arc tube and drops to zero near the wall. In this way, it is possible to account for the no-slip condition of the sound particle velocity~\cite{Kirchhoff.1868, Schuster.1940}.

\item Once the sound particle velocity is available, the force term driving AS can be determined. The AS~force is added to the buoyancy force term in the Navier-Stokes equation and the complete procedure has to be repeated starting from step 1. Thus, a recursion loop is initiated. Once convergence has been reached, the recursion is terminated.

\item The procedure is repeated for a slightly shifted frequency until the frequency range of interest is covered (frequency scan). In the simulations, the converged solution of the previous frequency is used as initial configuration for the simulation at the new frequency. This guarantees that the qualitative character of the solutions corresponding to neighboring frequencies is identical. Loss of convergence can then be interpreted as an indication that the respective solution does not represent a physically stable state anymore\footnote{This interpretation is supported by the results obtained by the accompanying experiments.}.

\end{enumerate}

Using the mode expansion method, despite the necessity of the recursion, leads to results within manageable computing times. The advantage is due to the fact that the acoustic model, obtained as described above, is linear. All details of the mode expansion approach can be found in \cite{BaumannEtAl.2015}.

\section{Results}
\begin{figure}
\centering
\includegraphics[width=0.45\linewidth]{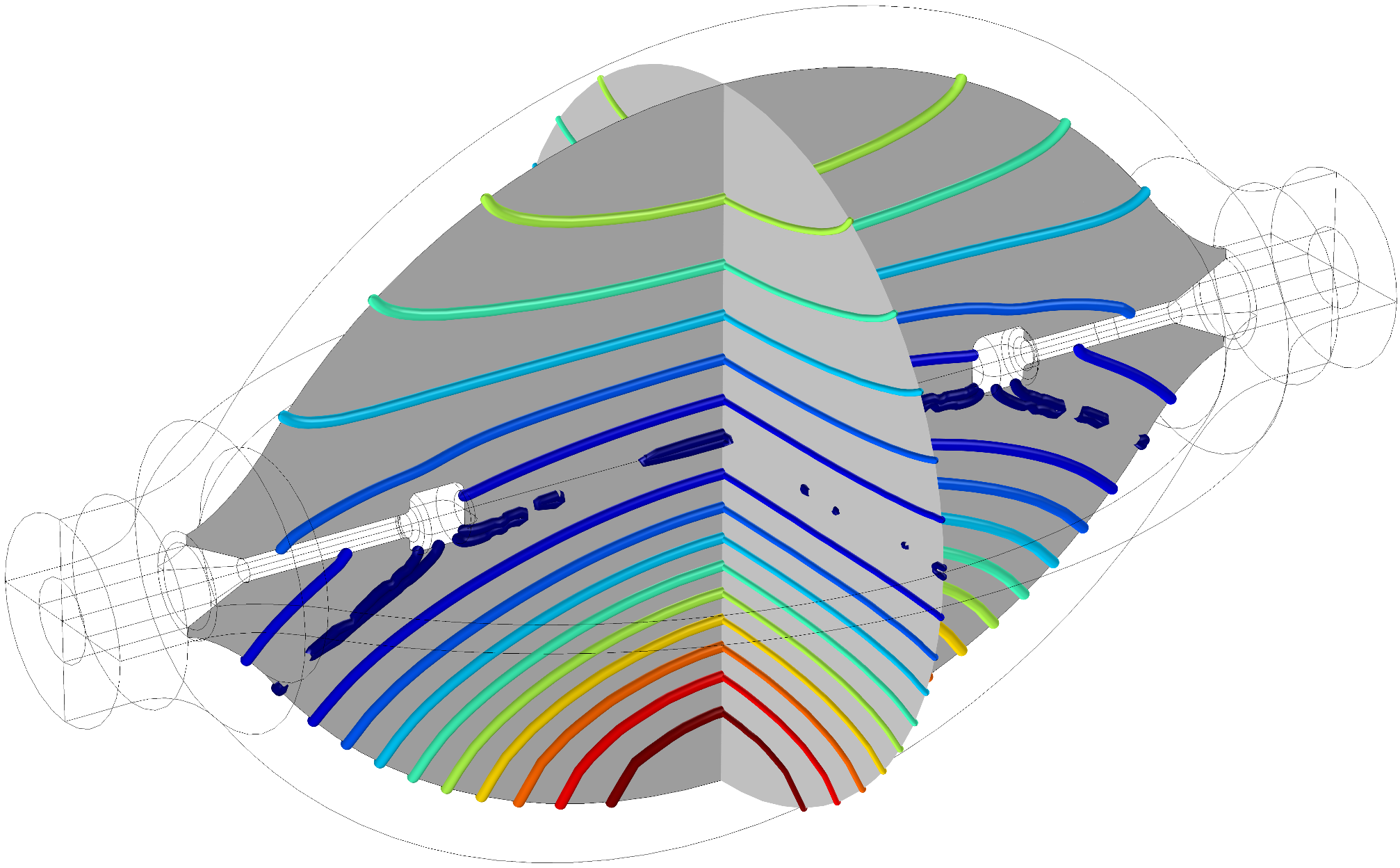}
\caption{Acoustic mode resulting from the first iteration of the recursion loop. The natural frequency is $\ze{47.8}{kHz}$. Depicted is the absolute value of the acoustic pressure in arbitrary units. Blue indicates the pressure node. The mode corresponds to the first iteration of the recursion loop (no AS~force).}
\label{fig:secondeigenmodetubes}
\end{figure}

For the lamps supplied by Philips, the lowest light flicker frequencies typically is near $\ze{42}{kHz}$. The FE~model identifies two acoustic modes in the vicinity of this frequency. The two modes are degenerate, if gravity and, therefore, the buoyancy force are switched off. In the presence of gravity, only the mode depicted in Figure~\ref{fig:secondeigenmodetubes} is excited and generates sound waves. The depicted mode corresponds to the first iteration of the recursion loop, when the AS~force is zero and the acoustic model (the Helmholtz equation) is not yet coupled to the plasma model. Close to resonance (see Section~\ref{sec:Natural Frequencies}) a strong AS~flow exists when the recursion has converged. The impact of the AS~flow field is a shift of the natural frequency, but no alteration to the general pattern of the mode.

\begin{figure}
\centering
{\includegraphics[width=0.99\linewidth,trim=0 0 0 0,clip]{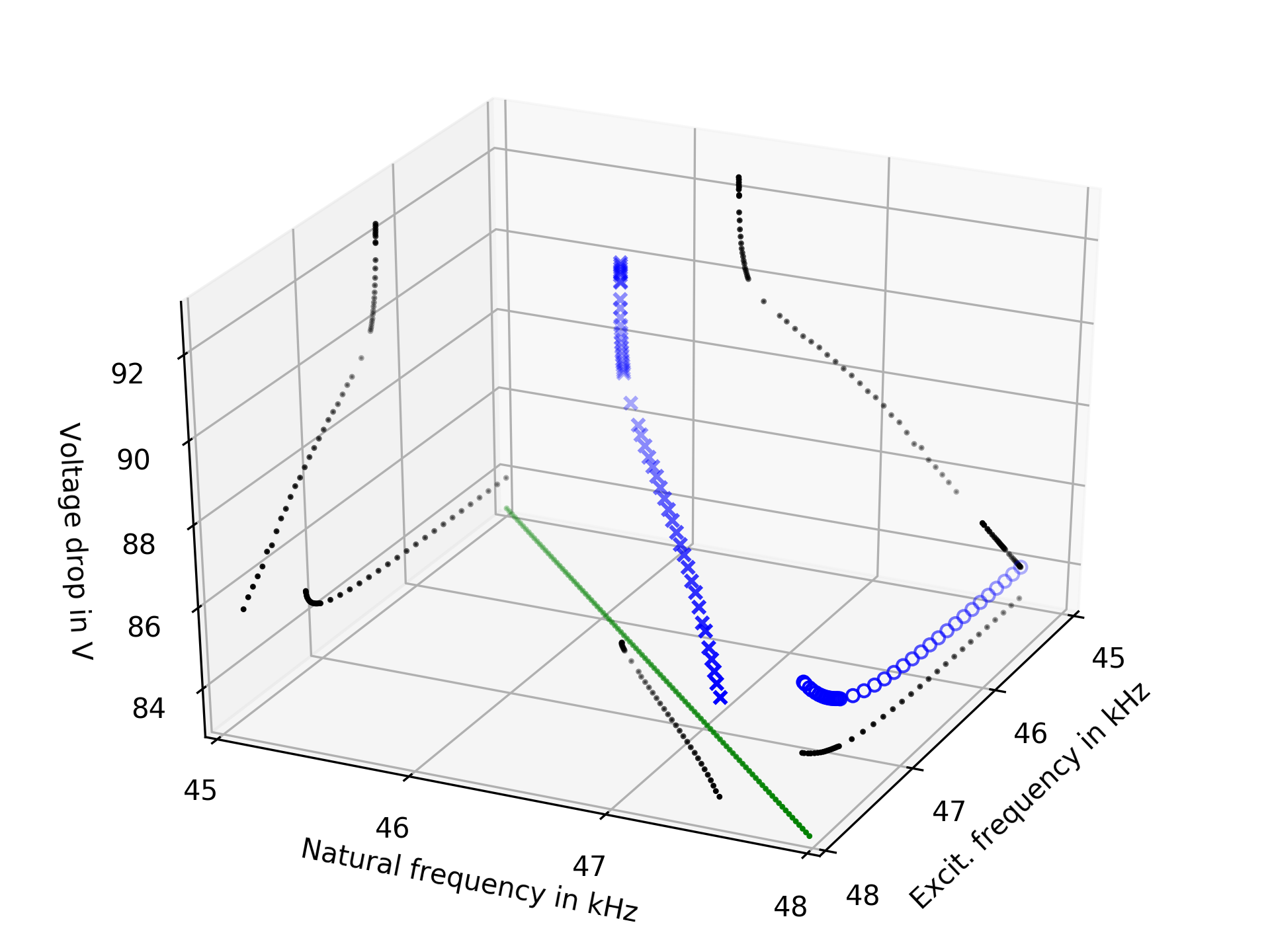}}
\caption{Connection of excitation frequency, natural frequency and voltage drop resulting from the FE~simulations (blue symbols). The black symbols result from projections of the blue curve to the respective plane. The meaning of the green line and of the distinction of data points represented by circles and crosses are explained in Section~\ref{sec:Natural Frequencies}.}
\label{fig:grafi3d1}
\end{figure}

In principle the interest is on the response of the acoustic pressure. The acoustic pressure data is available in the results of the FE~simulations, but there is no simple way for its measurement.
The quantity most easily accessible for measurement and also available from the data obtained in the FE~simulations is the voltage drop between the electrodes. Hence we focus on the voltage instead of the acoustic pressure. Figure~\ref{fig:grafi3d1} summarizes the corresponding simulation results. In the following we interpret the figure by discussing the projections of the curve depicted in blue to the three coordinate planes. During the discussion we will present evidence that the HID~lamp behaves in some respects similar to a nonlinear oscillator. As a prerequisite it seems helpful to recount some basic facts on oscillators.

An unforced, undamped oscillator with a linear spring oscillates with its natural frequency, which is determined by the stiffness of the spring and the mass of the bob. For a nonlinear spring, the stiffness varies with the displacement and, since the stiffness determines the natural frequency, the natural frequency depends on the amplitude of the oscillation. 

The bob of a forced oscillator moves with an amplitude that depends on the deviation of the forcing or excitation frequency and natural frequency. Therefore, for a nonlinear oscillator a shifting of the excitation frequency results in a variation of the natural frequency. Since the physical behavior of the HID~lamp is highly nonlinear, a corresponding behavior is expected and this assumption is confirmed in the following section.

\subsection{Natural Frequencies}
\label{sec:Natural Frequencies}
\begin{figure}
\centering
{\includegraphics[width=0.99\linewidth]{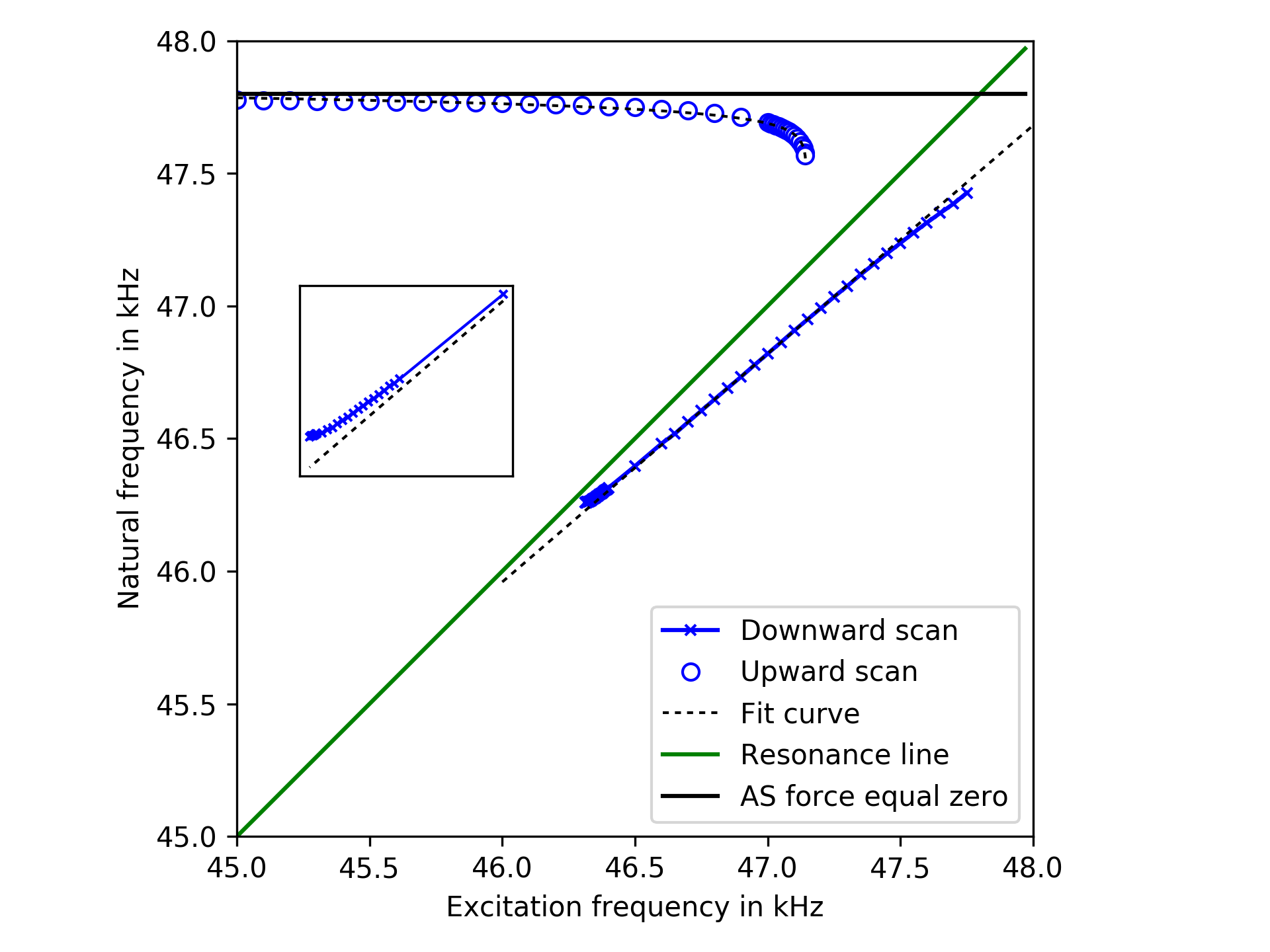}}
\caption{Natural frequency vs.~excitation frequency resulting from the FE~simulations. The inlay shows an enlargement of the downward scan data for $f_{\rm exc}\le\ze{46.50}{kHz}$.}
\label{fig:excfreqvseigenfreq1}
\end{figure}

Figure~\ref{fig:excfreqvseigenfreq1} shows how the natural frequency varies with the excitation frequency. Assuming weak damping, natural frequency and resonance frequency practically coincide. Therefore, when the excitation frequency is tuned to the natural frequency resonance occurs. Resonance corresponds to a point on the diagonal of Figure~\ref{fig:excfreqvseigenfreq1}.

When approaching resonance by an upward scan of the excitation frequency $f_{\rm exc}$, the natural frequency has an almost constant value just below the natural frequency of the mode of the linear acoustic system depicted in Figure~\ref{fig:secondeigenmodetubes}. This is to be expected, since far off the resonance the amplitude of the sound wave is small and the influence of the nonlinearity can be neglected. Only when the excitation frequency is near the resonance line, the nonlinearity shows in the data by a small drop of the natural frequency. The simulations cannot be extended to higher frequencies, since the simulations do not converge anymore. From the lack of convergence we conclude that near the resonances the system is unstable. We estimated the resonance frequency $f_{\rm res}$ by a fit of the data to the function $\; a\log(f_{\rm res}-f_{\rm exc}) + b\;$ and found $f_{\rm res}=\ze{47.15}{kHz}$. The values of the parameters $a$ and $b$ are not important here.  Figure~\ref{fig:excfreqvseigenfreq1} shows that the fit to the simulation results is excellent.

When approaching the resonance by a downward scan of the excitation frequency, the natural frequency drops by ca.~$\ze{1}{kHz}$. In this case it is possible to get very close to the resonance line before convergence gets lost. As expected, the data reveals that a variation of the excitation frequency results in a sizable shift of the natural frequency. At a first glance it appears as if the drop of the natural frequency in the downward scan is described by a straight line. A close inspection of the inlay in the figure reveals that the curve near the resonance systematically bends towards the resonance line. It is not clear to us which function should be used to describe the data. Therefore, we tried to estimate the corresponding resonance frequency by visual inspection. Our estimate is that it lies in the range between $\ze{46.25}{kHz}$ and $\ze{46.27}{kHz}$.

There is a second reason, why the description of the downward scan data by a straight line cannot be correct: As mentioned before, far off the resonance the influence of the nonlinearity can be neglected.
Therefore, an extrapolation of the data to higher excitation frequencies should result in an asymptotic approach to the black horizontal line representing the natural frequency of the linear acoustic model. 
Figure~\ref{fig:excfreqvseigenfreq1} depicts a straight line fit to the downward scan data for the excitation frequency range $\ze{46.65}{kHz}$ to $\ze{47.20}{kHz}$ in the middle of the complete downward scan frequency range. The deviation of the fit line and the data for $f_{\rm exc}> \ze{47.20}{kHz}$ indicates the beginning of the asymptotic approach.

\subsection{Backbone Curves}
\label{sec:Backbone Curves}
Next, we consider the plane spanned by the axis of the natural frequency and the axis of the voltage drop. As mentioned above the natural frequency of a nonlinear oscillator varies with the amplitude of the unforced oscillation. A graph which depicts the connection of natural frequency and amplitude is called \emph{backbone curve}\footnote{In a previous publication we used a different definition of the term \emph{backbone curve} \cite{0022-3727-49-25-255201}. In the leading order of a perturbation expansion the two definitions lead to identical expressions.} \cite{rand2012lecture}.
The backbone curve of a linear oscillator is trivial: Since the natural frequency constant, it is a straight line parallel to the amplitude axis. 

Almost 100 years ago, Georg Duffing investigated a certain type of idealized nonlinear oscillator~\cite{duffing1918erzwungene}. The restoring force of the \emph{Duffing oscillator} of angular frequency \nolinebreak{$\omega_0=1$} is described by
\nfm{F(x)=x+\gamma x^3,}
 where $x$ is the deflection of the bob from the position of equilibrium and $\gamma$ is a constants. For $\gamma>0$ the spring stiffens with increasing deflection and for $\gamma<0$ the spring becomes softer when it is stretched or compressed. If in the softening case the amplitude exceeds a critical value $x_{\rm crit}$, the force $F(x)$ changes its character from attractive ($F(x)<0$) to repulsive ($F(x)>0$), i.~e.~experiences a zero-crossing\footnote{A practical example for this is a ship floating in the water. If the ship is heeled by an external influence and then released, the righting moment will result in an oscillation of the ship. When the angle of heel increases to large values, the righting moment becomes weaker. At the critical angle of heel the righting moment changes its sign and the ship will capsize.}. Formally, $F(x)\le 0$  corresponds to a vanishing natural frequency.

\begin{figure}
\centering
\includegraphics[width=0.49\linewidth]{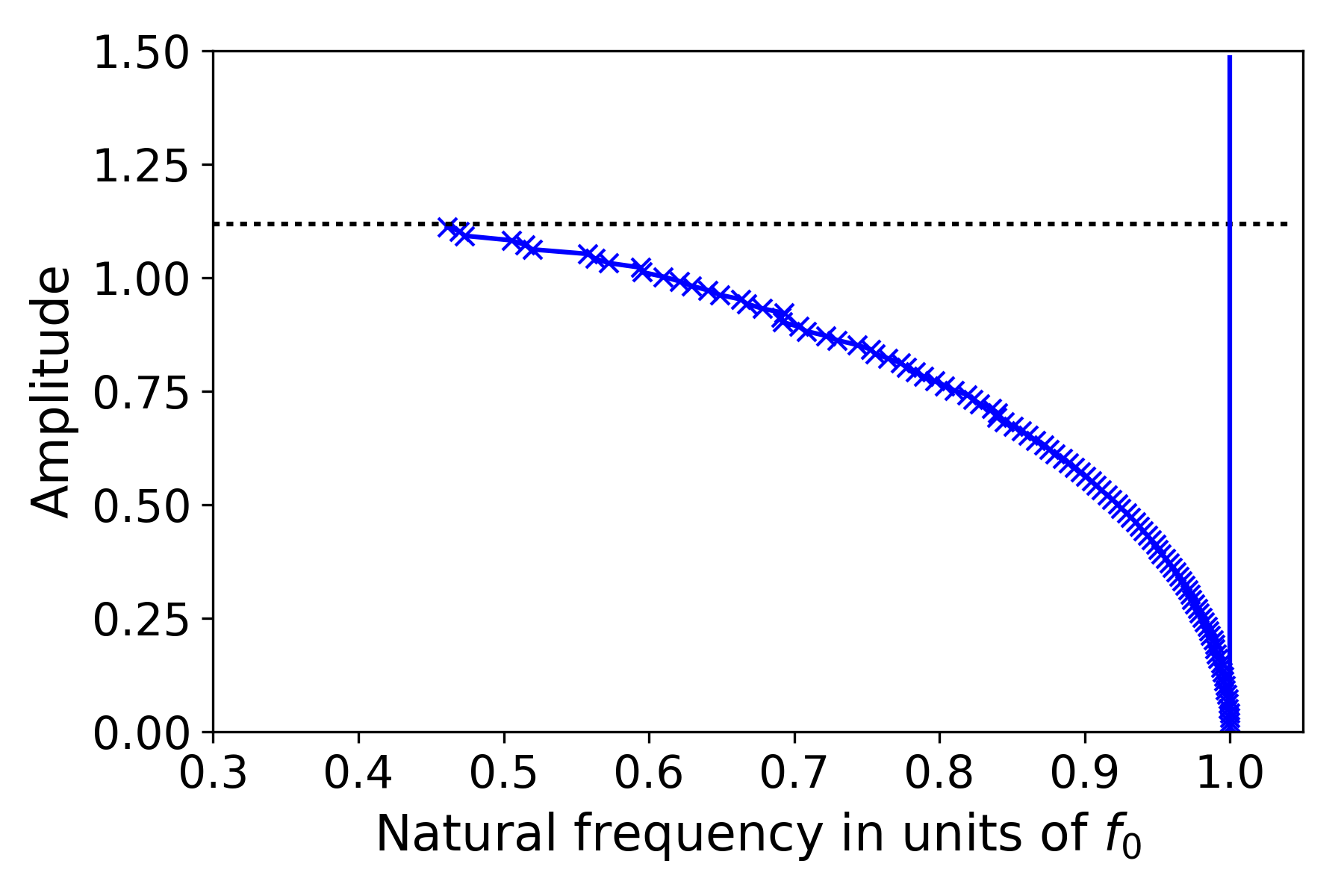}
\includegraphics[width=0.49\linewidth]{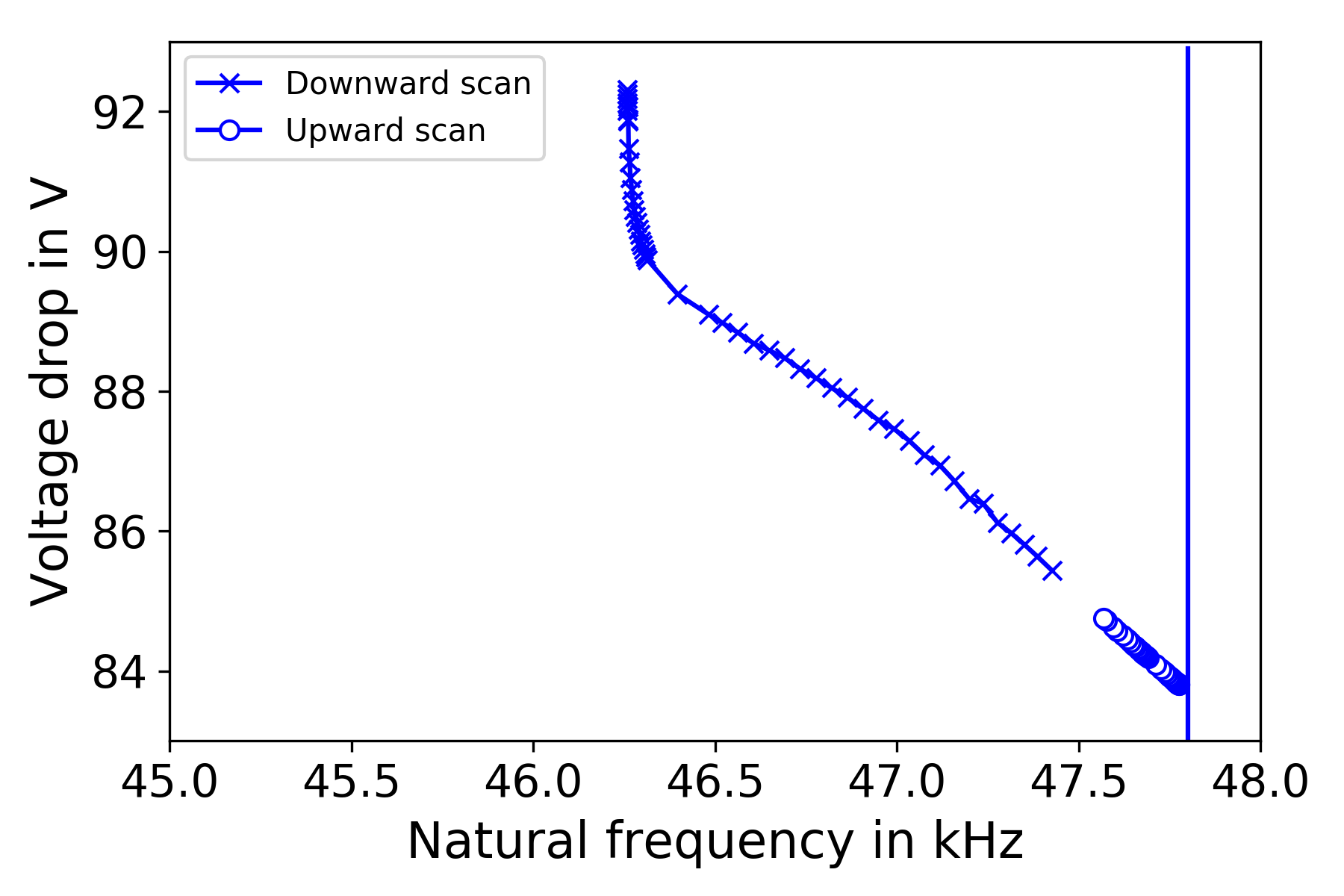}
\caption{Left: Backbone curve of a Duffing oscillator with a softening spring. The vertical blue line is the (trivial) backbone curve of the corresponding linear oscillator ($\gamma=0$). The dotted horizontal line corresponds to the critical amplitude $x_{\rm crit}$.  Right: Backbone curve resulting from the simulations of the HID~lamp. The vertical blue line is the (trivial) backbone curve resulting for AS~force equal to zero.}
\label{fig:backbone1}
\end{figure}

The left part of Figure~\ref{fig:backbone1} depicts the backbone curve of a Duffing oscillator with a softening spring. The curve has been obtained by solving the differential equation
\nfm{\ddot{x}+F(x) = 0}
numerically for $\gamma=-0.8$.  The curve ends once the amplitude reaches the critical value $x_{\rm crit}$. The right part of the figure depicts the backbone curve obtained from the simulations of the HID~lamp. The backbone curves of the two systems show a similar general behavior. The most prominent difference is the steepening of the HID~backbone curve, which is not present in the case of the Duffing oscillator. The steepening at the low frequency end of the curve, indicating the presence of a stiffening effect, continues until the slope of the backbone curve is vertical. An inspection of the flow field reveals that the occurrence of steepening is connected to the dominance by AS over buoyancy.

In a still idealized but more realistic model of an oscillator build from a softening spring and a bob (\emph{modified Duffing oscillator}), the steepening can be understood in the following way: At large amplitudes of the unforced oscillations the spring is stretched so far that the spring degenerates to a wire, respectively, the spring is compressed until neighboring coils come into contact. In both cases the systems experiences a dramatic stiffening. The force  exerted on the bob at such large deflections is much larger than at smaller deflections. For the sake of simplicity it can be assumed that the stretching and the behavior for compression can be described in identical ways, i.~e., that $F(-x)=-F(x)$.

 The forces acting on the bob at large deflections are practically identical to the forces acting on an elastic ball reflected between two walls. The bob/the ball is accelerated near the points of reversal and moves freely (ball) or almost freely (bob) during the rest of their trajectory. As long as the materials can be considered to be elastic, the bouncing frequency can be increased by applying appropriate initial conditions. In the case of the bob this is linked to an additional stretching of the wire/a quenching of the contacting coils and, therefore, a larger amplitude of the oscillation. Concerning the backbone curve this results in a steepening until the slope is vertical. It seems natural to assume that corresponding effects emerge in the case of the lamp and account for the steepening in Figure~\ref{fig:backbone1}.

\subsection{Forced Oscillations}

\begin{figure}
\centering
\includegraphics[width=0.49\linewidth]{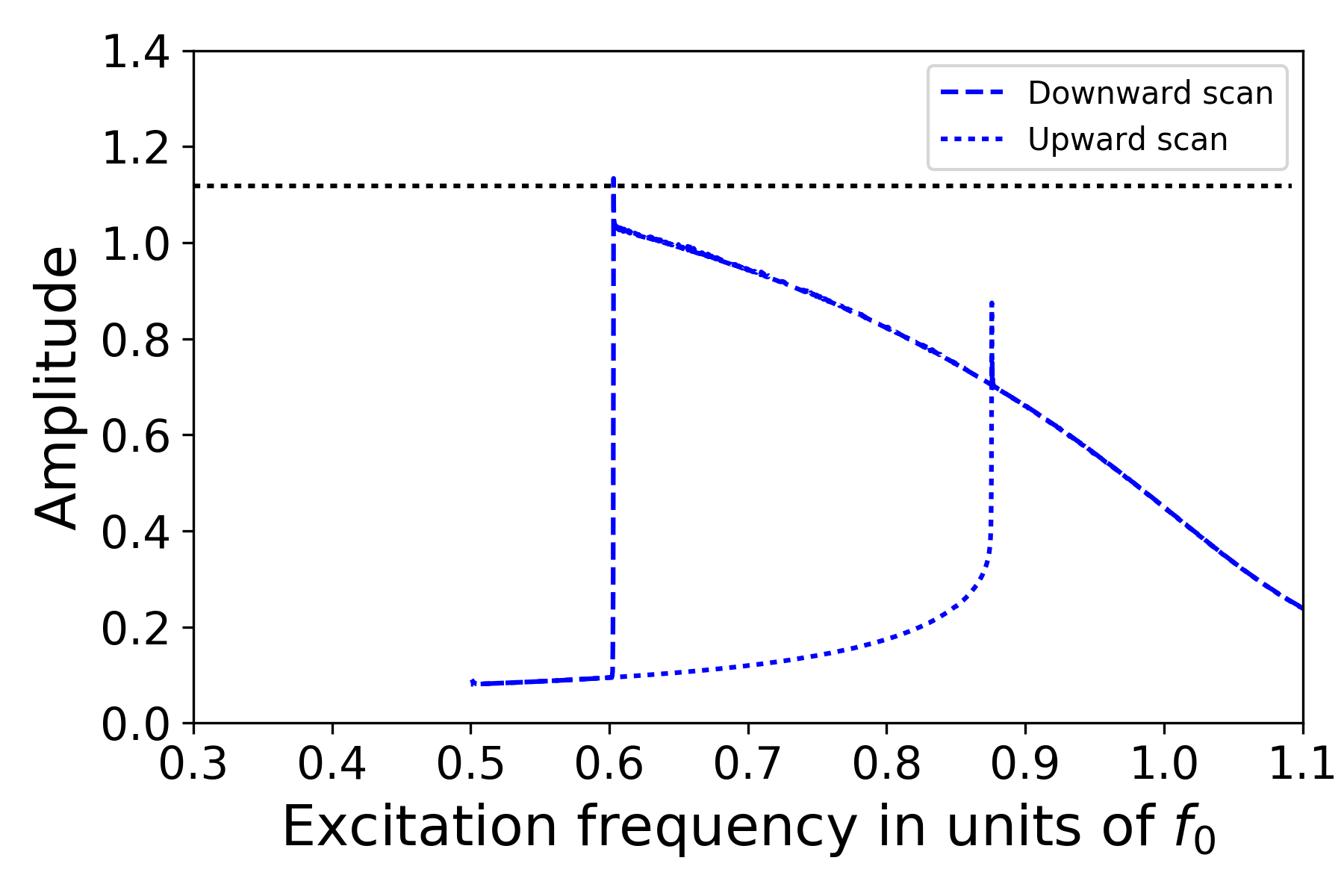}
\includegraphics[width=0.49\linewidth]{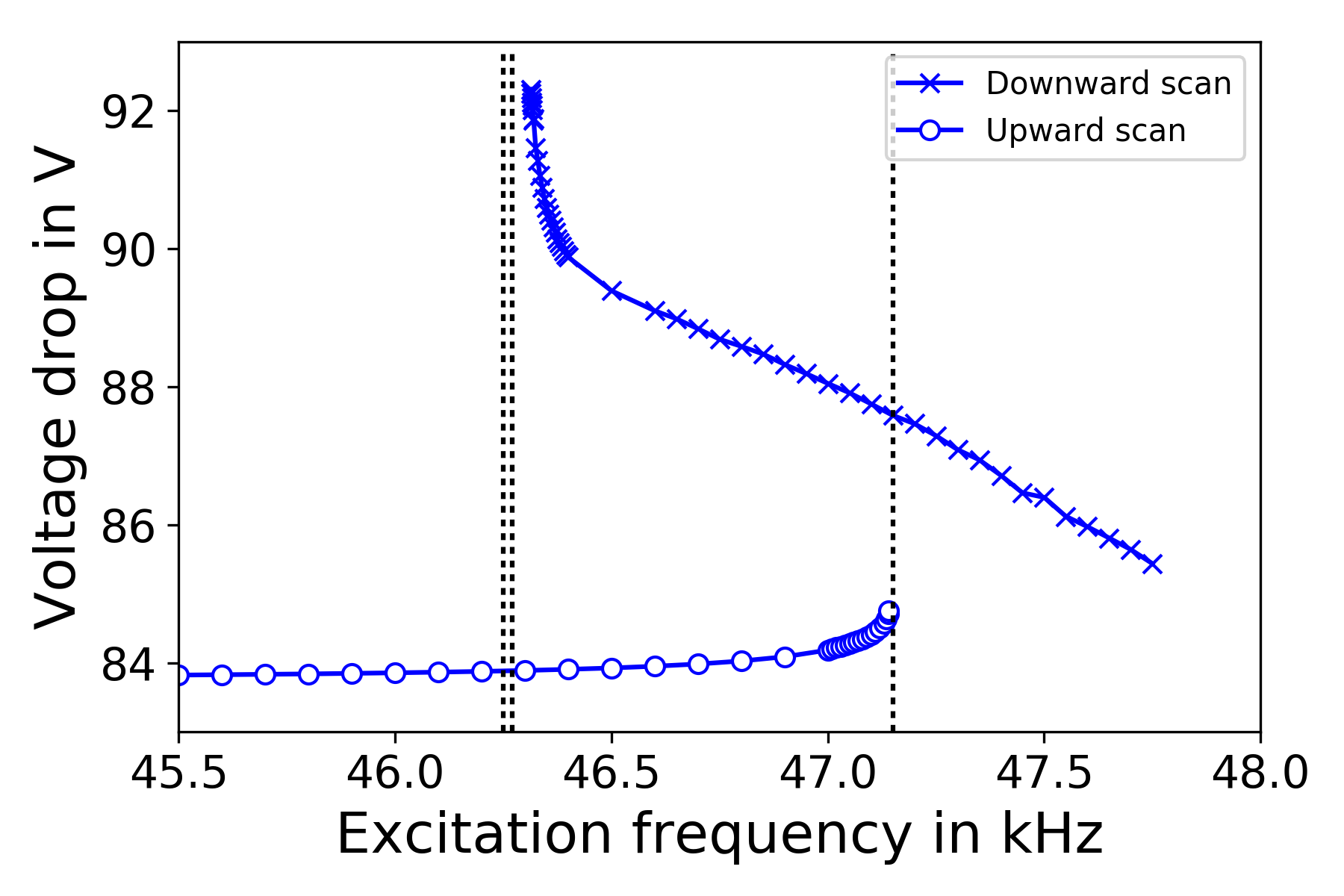}
\caption{Left: Response of a Duffing oscillator with a softening spring. The dotted horizontal line corresponds to the critical amplitude. Outside the hysteresis region the up- and downward data are practically identical and, therefore, not discernible. At the jump frequencies the response shows overshoots, which are related to the dynamics of the jumps. Right: Results from the simulations of the HID~lamp for the response of the voltage drop. The dotted vertical lines indicate the location of the resonance frequencies (Section \ref{sec:Natural Frequencies}).}
\label{fig:Response1}
\end{figure}

Finally we consider the plane spanned by the axis of the excitation frequency and the axis of the voltage drop. The left part of Figure~\ref{fig:Response1} shows the response of a Duffing oscillator with a softening spring. The frequency scans have been calculated by numerically solving the differential  equation
\nfm{\ddot{x}+2\zeta\dot{x}+F(x)= x_0\cos(\Omega\tau)}
with $\gamma=-0.8$, $\zeta=0.028$ and $x_0=0.06$ and by varying the angular frequency $\Omega$ in small steps. In the upward scan the system jumps at $\Omega_1=0.875$ to a considerably larger value of the amplitude and in the downward scan to a smaller amplitude value at $\Omega_2=0.602$, a behavior well known from the Duffing oscillator. The two frequencies $f_1:=\Omega_1/2\pi$ and $f_2:=\Omega_2/2\pi$ are called \emph{jump frequencies}. The different behaviors of the system in up- and downward frequency scans exhibit the characteristic of a hysteresis.

The right part of Figure~\ref{fig:Response1} shows the results for the response of the voltage drop obtained from the simulations of the HID~lamp. Again, strong similarities in the behavior of the Duffing oscillator and the HID~lamp can be detected. The jump phenomenon offers an explanation for the light flicker effect.
Obviously, no data from the upward frequency scan exists above the resonance frequency resulting from the fit in Section \ref{sec:Natural Frequencies}. The estimated boundaries of the range of resonance frequencies for the downward scan are also shown in the figure. Similarly to the backbone curves, the upper branches of the response curves for the Duffing oscillator and the voltage drop exhibit a pronounced distinction: Near the low frequency end the voltage drop curve steepens strongly, while the response of the Duffing oscillator does not show such a behavior. It is obvious that this steepening is related to the stiffening effect described in Section \ref{sec:Backbone Curves}. In the experimentally measured voltage data a very similar behavior compared to the one depicted in the right part of Figure~\ref{fig:Response1} has been found \cite{0022-3727-49-25-255201}.

\begin{figure}
\centering
\includegraphics[width=0.49\linewidth]{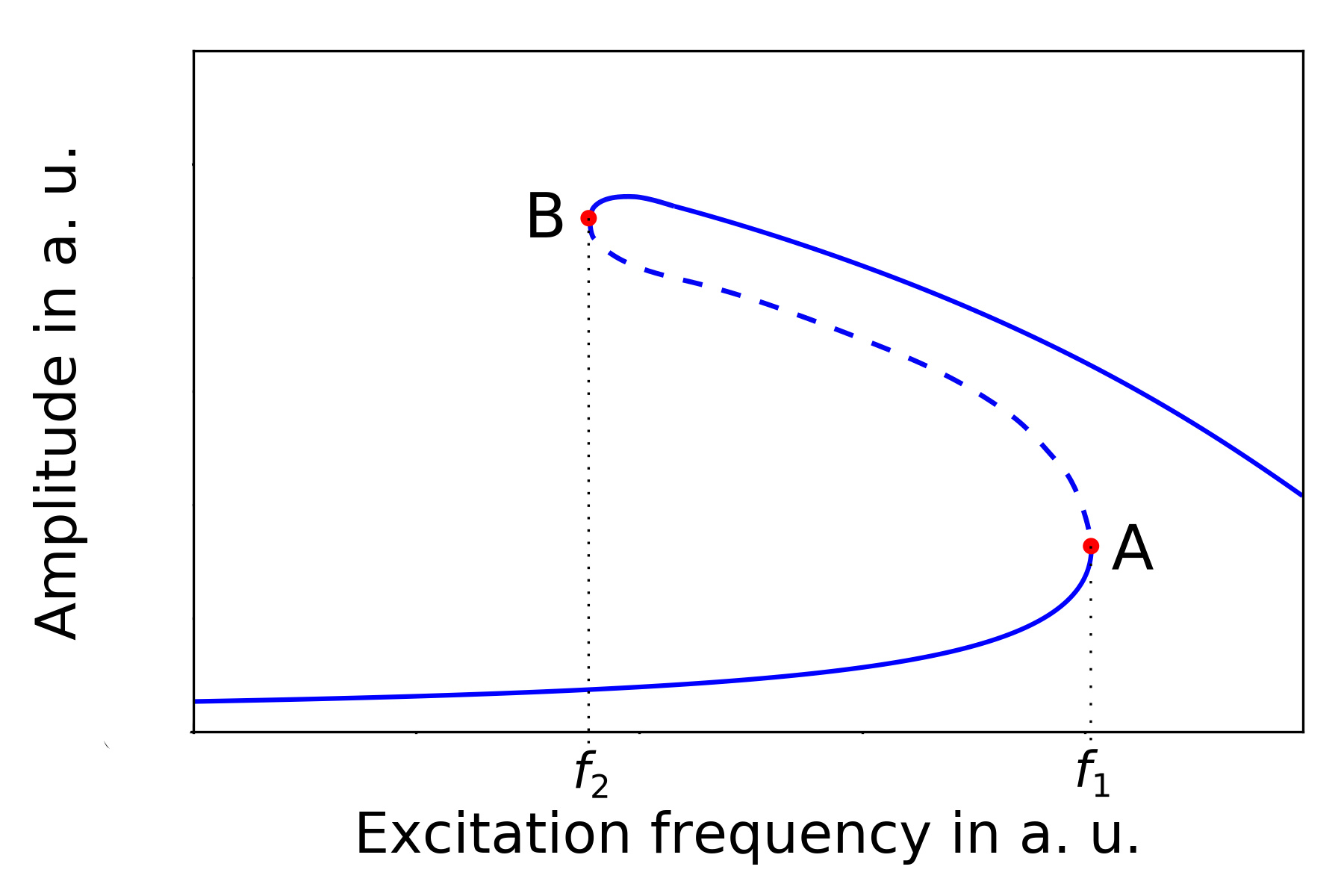}
\newline
\includegraphics[width=0.49\linewidth]{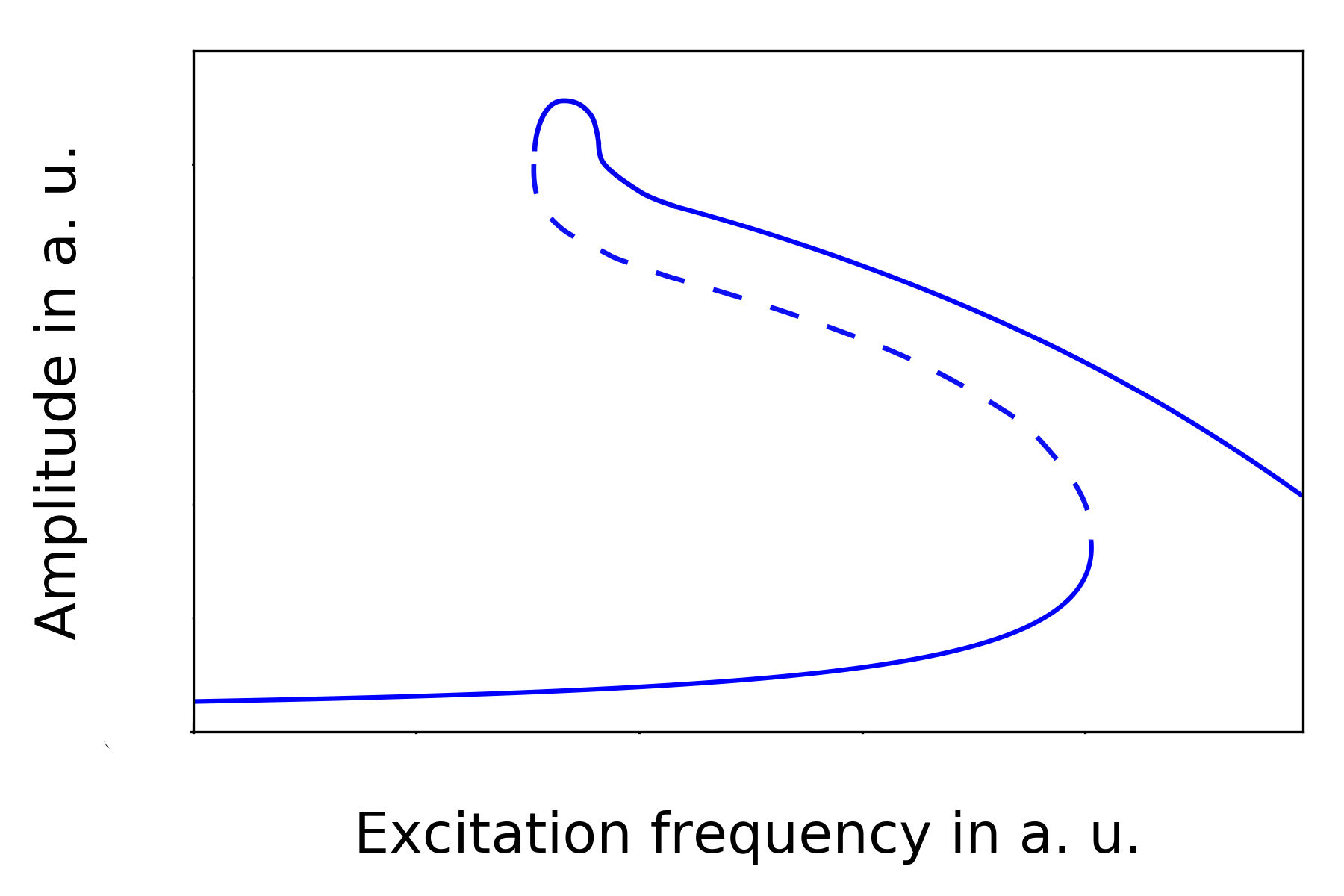}
\newline
\includegraphics[width=0.49\linewidth]{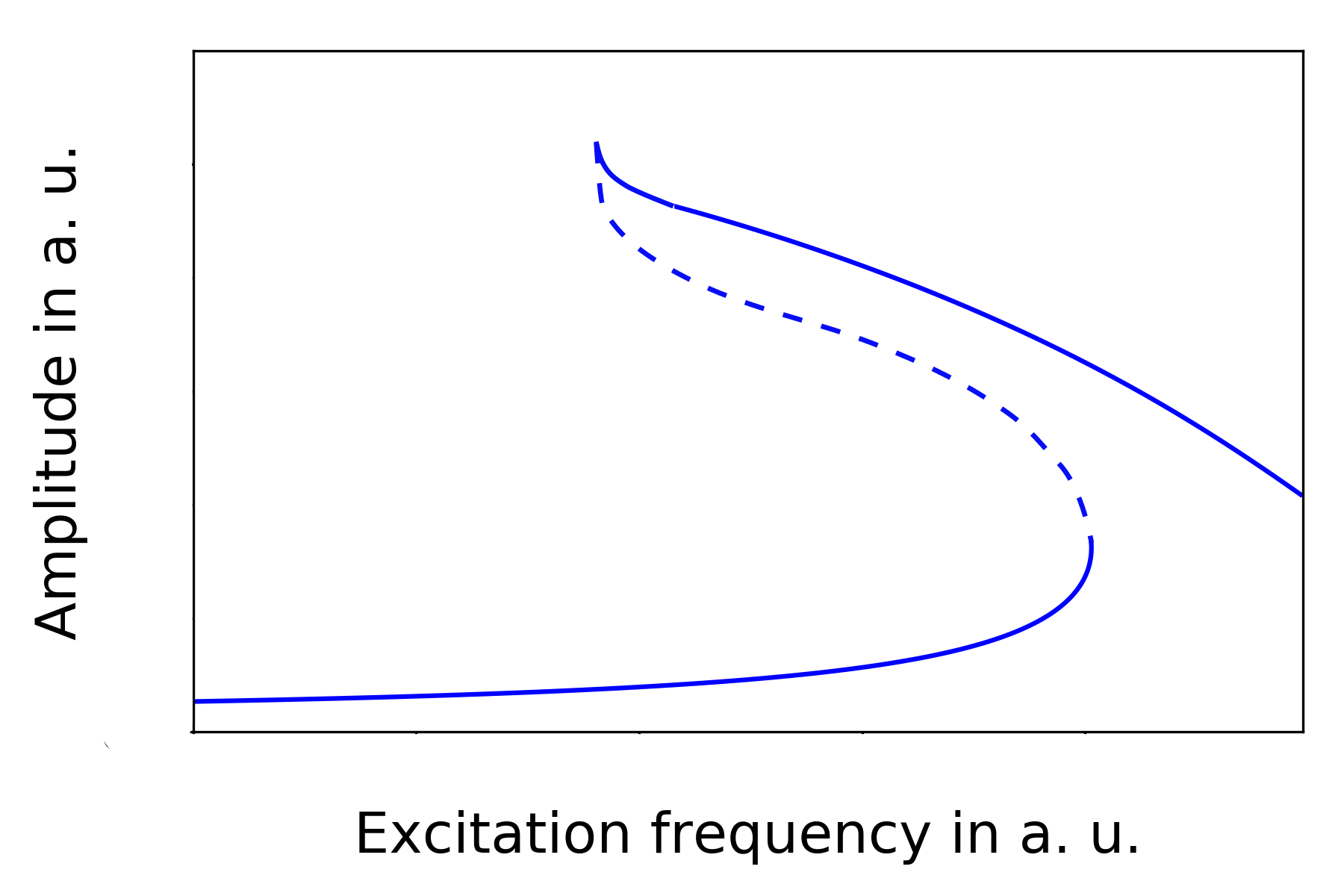}
\newline
\caption{Top: Response of the Duffing oscillator with softening spring \cite{kalmar2011forced}. The dashed part of the curve corresponds to an unstable solution. Middle and bottom: Tentative suggestions of response curves for the modified Duffing oscillator, respectively, the HID lamp.}
\label{fig:Response2}
\end{figure}

The top part of Figure~\ref{fig:Response2} qualitatively depicts the response curve of a Duffing oscillator with a softening spring, which is obtained by applying a perturbation expansion \cite{kalmar2011forced}.
 All frequencies below $f_2$ and above $f_1$ correspond to a unique value of the amplitude. However, in the range from above $f_2$ to below $f_1$ the amplitude can attain three different values. The branch of the response ranging from point {\sffamily A} to point {\sffamily B} represents an unstable solution. Therefore, the frequency range from $f_2$ to $f_1$ is called \emph{interval of bi-stability}. It is in the nature of the process that the unstable branch of the solution does not show in the numerical results depicted in Figure~\ref{fig:Response1}. 

At the jump points {\sffamily A} and {\sffamily B} the tangent to the response curve is vertical. Each of these two points constitutes a bifurcation point of a \emph{cyclic fold bifurcation}~\cite{thomsen2002some}. It would be interesting to identify the type of the bifurcation at the jump points of the response of the HID lamp. However, the determination of the the system's bifurcation type would require to analyze the eigenvalues of the Jacobian and we are not able to do that.


In the following we assume that the two stable branches of the response curve are connected by a branch representing an unstable solution. Finding a way to connect the stable branches of the response of the HID lamp, this is not possible without complications due to the steepening. If we rule out a scenario with crossing branches, we only have two choices:

\begin{itemize}
\item The most obvious way is schematically depicted in the middle part of Figure~\ref{fig:Response2}. This scenarios possesses one additional point of the response curve with a vertical slope. For the HID lamp behavior we rule out this scenario, because it should show in the simulation.


\item The second possibility to connect the two stable branches is depicted in the bottom part of Figure~\ref{fig:Response2}. In this case the response curve is not smooth. Instead it has a singularity at the point of bifurcation. This seems to be the most likely scenario. 
\end{itemize}

To our knowledge, no qualitatively different scenarios offer a consistent description of the observed behavior so that no other consequences as the described ones are possible for the behavior of the corresponding physical system. Unless we drop the assumption that the stable branches of the response are connected by an unstable branch and the unstable branch must not cross a stable branch, we do not see a way to avoid such a scenario.

\section{Conclusions}
A stationary 3D FE~model for the simulation of the processes in the arc tube of a certain type of HID~lamps has been set up.
The aim was to obtain a better understanding of the light flicker phenomenon, which worries the lighting industry for decades.
In order to restrict the necessary computer resources, the model includes a simplified description of the plasma and the acoustic streaming effect. 
The results of the simulations reveal that the response of the driven system exhibits properties, which strongly resemble that of the forced Duffing oscillator with a softening spring. In particular,
a hysteresis and the jump phenomenon have been observed.
However, the HID lamp exhibits features, which do not have a counterpart in the Duffing oscillator. It seems that it is necessary to use an new type of response curve to describe the behavior of the HID lamp properly.

\vspace{3mm}
{\noindent\bf Acknowledgment:} 
This research was supported by the German Federal Ministry of Education and Research (BMBF) under project reference 03FH025PX2 and Philips Lighting. 

\pagebreak
\bibliographystyle{unsrt}
\bibliography{/Users/BB/Desktop/Dokumente/Forschung/HID-Lamp/Paper/HIDLiteratur}   

\end{document}